\documentclass[12pt]{article}
\usepackage{epsfig}

\usepackage{rotating}

\usepackage{epsfig}
\usepackage{bm}

\newcommand{\be}{\begin{equation}}
\newcommand{\ee}{\end{equation}}

\newcommand{\bea}{\begin{eqnarray}}
\newcommand{\eea}{\end{eqnarray}}

\title{Perturbative QCD analysis of the Bjorken sum rule
\footnote{The work was supported by RFBR grant No.10-02-01259-a}}
\author{  A.V. Kotikov and B.G. Shaikhatdenov\\
Joint Institute for Nuclear Research, Russia \\
}

\begin{document}
\maketitle


\abstract{
We study the polarized Bjorken sum rule at low momentum transfer squared
$Q^2 \leq 3$ GeV$^2$ in the twist-two approximation and to
the next-to-next-to-leading order accuracy.
}



\section{Introduction}
The spin structure of a nucleon is one of the most interesting problems
to be resolved within the framework of (nonperturbative) Quantum Chromodynamics (QCD).
In particular, the singlet part $\Sigma(x,Q^2)$ of the parton distribution
functions
$$\Sigma(x,Q^2)=\sum_{i=1}^f f_a(x,Q^2),$$
where $f$ is a number of active quarks, is intensively studied,
because there is strong disagreement between the experimental data
for its first Mellin moment
and corresponding theoretical predictions. This disagreement is usually
called a spin crisis (see, for example, reviews in
\cite{Anselmino:1994gn}).

Here we consider only
the non-singlet (NS) part, which
the  fundamental Bjorken sum rule (BSR) holds for~\cite{Bjorken:1966jh}
$$
\Gamma^{p-n}_1(Q^2) = \int^1_0 \Bigl[g_1^p(x,Q^2)-g_1^n(x,Q^2)\Bigr] dx.
$$
It deals with the first moment ($n=1$) of NS part of the structure function (SF)
$g_1(x,Q^2)$.
For the case $n=1$,
the corresponding anomalous dimension
of Wilson operators is zero
and all the $Q^2$-dependence
of $\Gamma^{p-n}_1(Q^2)$ is encoded in the
coefficient function.

Usually, BSR is represented in the form
$$
\Gamma^{p-n}_1(Q^2) = \frac{g_A}{6} E_{NS}(Q^2) + \sum_{i=2}^{\infty}
\frac{\mu_{2i}^{p-n}(Q^2)}{Q^{2i-2}},
$$
where the first term in the r.h.s. is a twist-two
part and the second one is
a contribution of higher twists (HTs).

At high $Q^2$ values the experiment data for $\Gamma^{p-n}_1(Q^2)$ and the theoretical predictions
\cite{Anselmino:1994gn} are well compatible with each other.
Here we will focus on low $Q^2$ values, at which there presently exist the very precise
CLAS
\cite{Dharmawardane:2006zd,Amarian:2002ar} and
SLAC \cite{Abe:1997cx} experimental data for BSR.
On the other hand, there also is a great progress in theoretical calculations:
recently, the terms $\sim \alpha_s^4$ are evaluated in \cite{Baikov:2010je}.

\begin{figure}[t]
\centering
\vskip 0.5cm
\includegraphics[width=0.6\textheight]{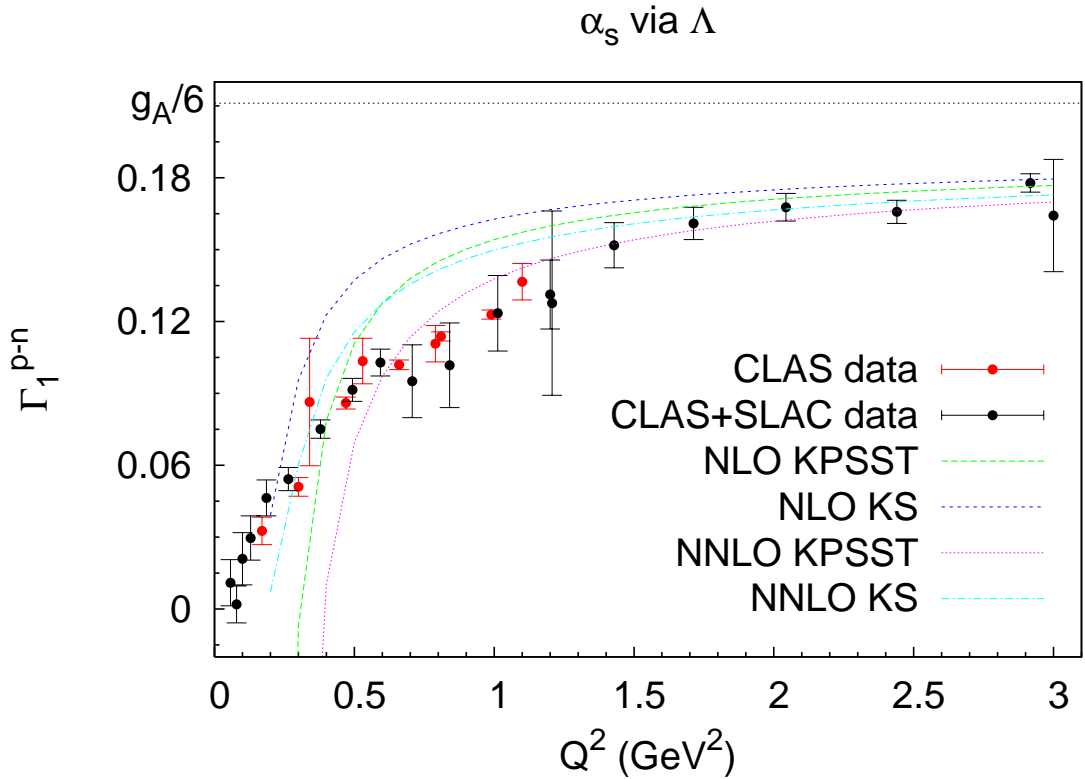}
\vskip -0.3cm
\caption{(color online).
CLAS \cite{Dharmawardane:2006zd,Amarian:2002ar} and
SLAC \cite{Abe:1997cx} experimental data for BSR and $Q^2 \leq 3$ GeV$^2$.
The curves represent theoretical predictions obtained in the analyzes carried out by
two groups: Khandramai, Pasechnik, Shirkov, Solovtsova, and Teryaev (KPSST)~\cite{Khandramai:2011zd} and Kotikov and Shaikhatdenov (KS).}
\label{fig:F1}
\end{figure}

\section{Basic formulae
}

In our analysis we will mostly follow the analyses done by the Dubna-Gomel
group
\cite{Khandramai:2011zd,Pasechnik:2009yc}.
We try, however, to resum the twist-two part with the purpose of reducing
a contribution coming from the HT
terms.

Indeed, there is an interplay
\begin{itemize}
\item
between HTs and higher orders of perturbative QCD
corrections (see, for example, \cite {Kataev:1996vu},
where the SF $xF_3$ was analyzed).
\item
between HTs and resummations in the twist-two part (see, for example,
application of the Grunberg approach \cite{Grunberg:1980ja} in
\cite{Parente:1994bf} to the study of SFs $F_2$ and $F_L$)
\end{itemize}

The twist-two part of BSR has the following form (see, for example,
\cite{Khandramai:2011zd})
\be
 E_{NS}(Q^2) = 1 -4 \Delta(Q^2),
\label{1}
\ee
where the term $\Delta(Q^2)$ looks like
\be
\Delta(Q^2) = a_s(Q^2) \Bigl(1+ \sum_{k=1}^{\infty} C_k a^k_s(Q^2)\Bigr)
~~~ \left(a_s(Q^2) = \frac{\alpha_s(Q^2)}{4\pi} \right) \, .
\label{1a}
\ee

The first three coefficients $C_1$, $C_2$ and $C_3$ are already known
(see \cite{Baikov:2010je,Larin:1991tj}
and references therein).

We will replace the above representation (\ref{1a})
by the following one
\be
 E_{NS}(Q^2) = \frac{1}{1 +4 \tilde{\Delta}(Q^2)},
\label{2}
\ee
where
\be
\tilde{\Delta}(Q^2) = a_s(Q^2) \Bigl(1+ \sum_{k=1}^{\infty}
\tilde{C}_k a^k_s(Q^2)\Bigr)
\label{1b}
\ee
and $\tilde{C}_k$ can be obtained from the known $C_k$:
\be
\tilde{C}_1 = C_1+4,~~\tilde{C}_2 = C_2+8C_1+16,~~\tilde{C}_2 = C_3+8C_2
+4C_1^1+48C_1+64 \, .
\label{2a}
\ee

The reason behind this transformation is as follows:
the CLAS experimental data  \cite{Dharmawardane:2006zd,Amarian:2002ar}
demonstrate that $\Gamma^{p-n}_1(Q^2\to 0) \to 0$.
Therefore, in the case when the HT
corrections produce small contributions at $Q^2 \to 0$
\footnote{
It is for sure questionable, but in the
KPSST analysis
\cite{Khandramai:2011zd}
$\mu_4 \sim 0$ at the next-to-next-to-next-to-leading order (N$^3$LO)
accuracy.}
we see that
\be
  E_{NS}(Q^2\to 0) \to 0 \, .
\label{2b}
\ee

Since the strong coupling constant $a_s(Q^2\to \Lambda^2) \to \infty$,
it is seen that the form (\ref{2}) behaves much like
the CLAS experimenatal data. Indeed,
\be
E_{NS}(Q^2\to \Lambda^2) = \frac{1}{1 +4 \tilde{\Delta}(Q^2\to \Lambda^2)}
\to 0 \, .
\label{2c}
\ee

As
$\Lambda^2_{QCD} \sim 0.01$ is rather small, one can conclude
that the above representation (\ref{2c}) agrees with experiment at very
low $Q^2$ values.

Note, however, that we have a very small coefficients of
$\Delta(Q^2)$ and $\tilde{\Delta}(Q^2)$.
Thus, for small but nonzero $Q^2$
values
the above representations (1) and (2) lead to similar results
(see Fig. 1, where we restricted our consideration to the
next-to-next-to-leading order (NNLO) accuracy).
As is seen in Fig. 1, the theoretical predictions are not
too close to the shape of the experimental data.

\section{Grunberg approach}
At $Q^2 \sim 0$, the value of the strong coupling constant is very large.
Thus, in our approach it is better to avoid the usage of series like
\be
\sum_{k=1}^{\infty} C_k a^k_s(Q^2) \, ,
\label{3}
\ee
as in Eqs. (\ref{1a}) and (\ref{1b}).

Instead, it is convenient
to use the Grunberg method of effective charges
\cite{Grunberg:1980ja}, i.e. to consider the variables $\Delta(Q^2)$
and $\tilde{\Delta}(Q^2)$ as new effective
``coupling constants'' \footnote{A similar application can be found in
the recent paper \cite{Shirkov:2012ux}.}, which
have the following properties:
\begin{itemize}
\item
shifted arguments
\be
Q^2 \to Q^2/D_k,~~~ Q^2 \to Q^2/\tilde{D}_k
\label{4}
\ee
for the variables $\Delta(Q^2)$ and $\tilde{\Delta}(Q^2)$, respectively,
with
\be
D_k =e^{C_1/\beta_0},~~ \tilde{D}_k=e^{\tilde{C}_1/\beta_0},
\label{5}
\ee
which are in turn responsible for the vanishing of
the coefficients $C_1$ and $\tilde{C}_1$ in a series similar to (\ref{3}).
Moreover,
these shifted arguments (\ref{4}) provide also a strong reduction
in the magnitudes of the coefficients $C_k$ and $\tilde{C}_k$ ($k\geq 2$).

\item
new $\beta_i$ ($i\geq2$) coefficients of the corresponding $\beta$-functions,
which are responsible for the vanishing
of the coefficients $C_k$ and  $\tilde{C}_k$ ($k\geq 2$).
\end{itemize}

However, a straightforward application of the Grunberg approach to
the variables $\Delta(Q^2)$ and $\tilde{\Delta}(Q^2)$ is not as convenient,
because the coefficients $C_1$ and $\tilde{C}_1$ are positive and the
$Q^2$ values are very small. It is in contrast with its
direct applications, where
the coefficients $C_1$ and $\tilde{C}_1$ are negative \cite{Kotikov:1993yw}
and/or the $Q^2$ values are not so small \cite{Parente:1994bf,Kazakov:1987jk}.

So, the
new arguments $Q^2/D_k$ and $Q^2/\tilde{D}_k$ have now very small values and, as a result,
we have to use the Grunberg approach associated with something else.
One of the ways is to use a so-called ``frozen'' coupling constant.

\section{ ``Frozen'' coupling constant}

We introduce freezing
of the coupling constant by altering its argument $Q^2 \to
Q^2_a = Q^2 + a M^2_{\rho}$,
where $M_{\rho}$ is
a $\rho$-meson mass and $a$ is some free parameter
(usually, $a=1$ was used. See, for example, \cite{Badelek:1996ap}).

Thus, in the formulae of the previous
sections the following replacement should be done
(a list of references can be found in \cite{Kotikov:2004uf}):
\begin{equation}
 a_s(Q^2) \to a_{fr}(Q^2) \equiv a_s(Q^2 + a M^2_{\rho})
\label{Intro:2}
\end{equation}

In the analyzes given below we restrict ourselves to
the next-to-leading order (NLO) (i.e. $\alpha_s^2$) approximation.
The consideration of two even higher order corrections is in progress.

\begin{figure}[tb]
\centering
\vskip 0.5cm
\includegraphics[width=0.6\textheight]{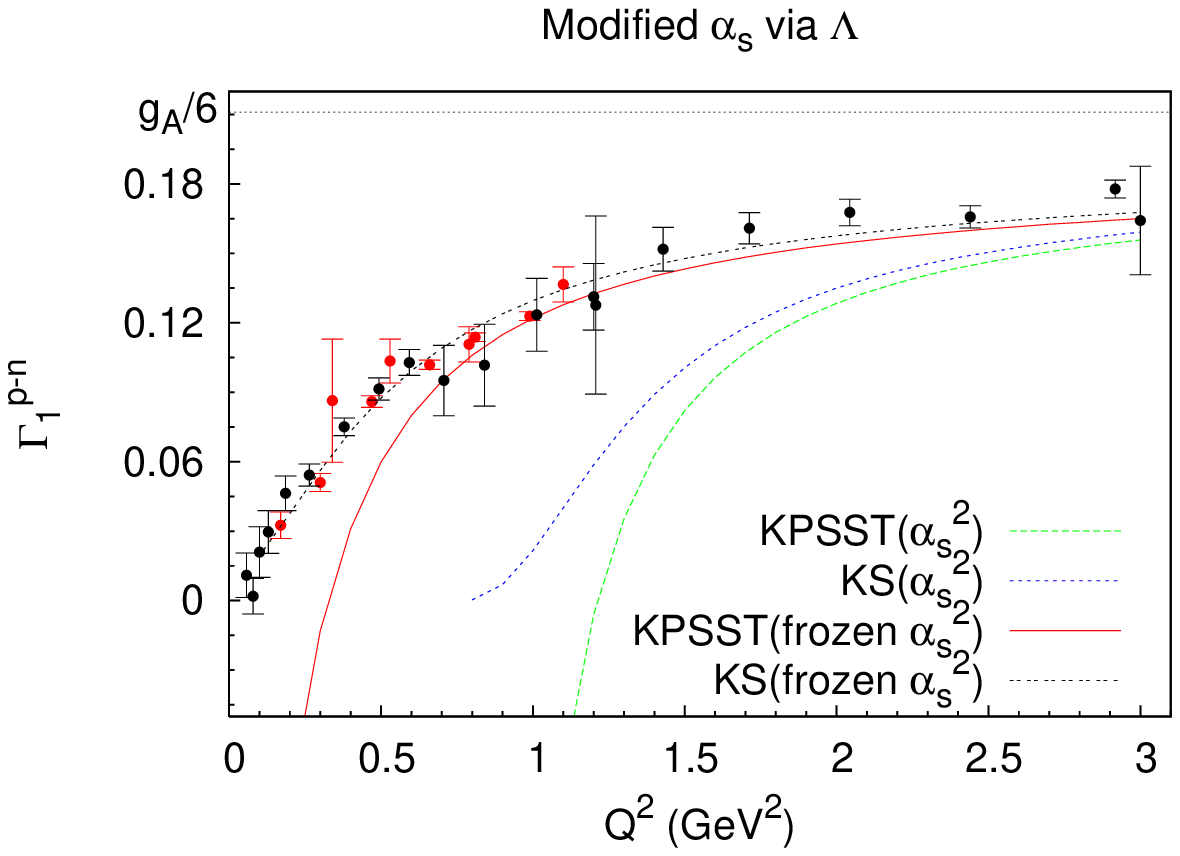}
\vskip -0.3cm
\caption{(color online). The experimental data are as in Fig.1.
The curves show theoretical predictions based on Eqs. (\ref{1})
and (\ref{2}), which is called a KPSST-like analysis \cite{Khandramai:2011zd}
and a KS one, respectively. For all cases the Grunberg approach
\cite{Grunberg:1980ja} is used with a standard coupling
constant and with a ``frozen'' one, when $a=1.5$.}
\label{fig:F2}
\end{figure}

The cases with
$a=1.5$ and $a=2$ are shown in Figs. 2 and 3, respectively.
It is seen that the best
agreement with experimental data is achieved
in the case of representation (\ref{2}) and $a=1.5$. Also,
the standard form (\ref{1}) and theoretical predictions
obtained with $a=2$ are well consistent with each other.

In contrast with the analyzes carried out with a standard coupling constant (see Fig. 1),
we observe that the shape of theoretical
predictions and the form of experimental data are close enough to each other at small
$Q^2$ values.

\begin{figure}[tb]
\centering
\vskip 0.5cm
\includegraphics[width=0.6\textheight]{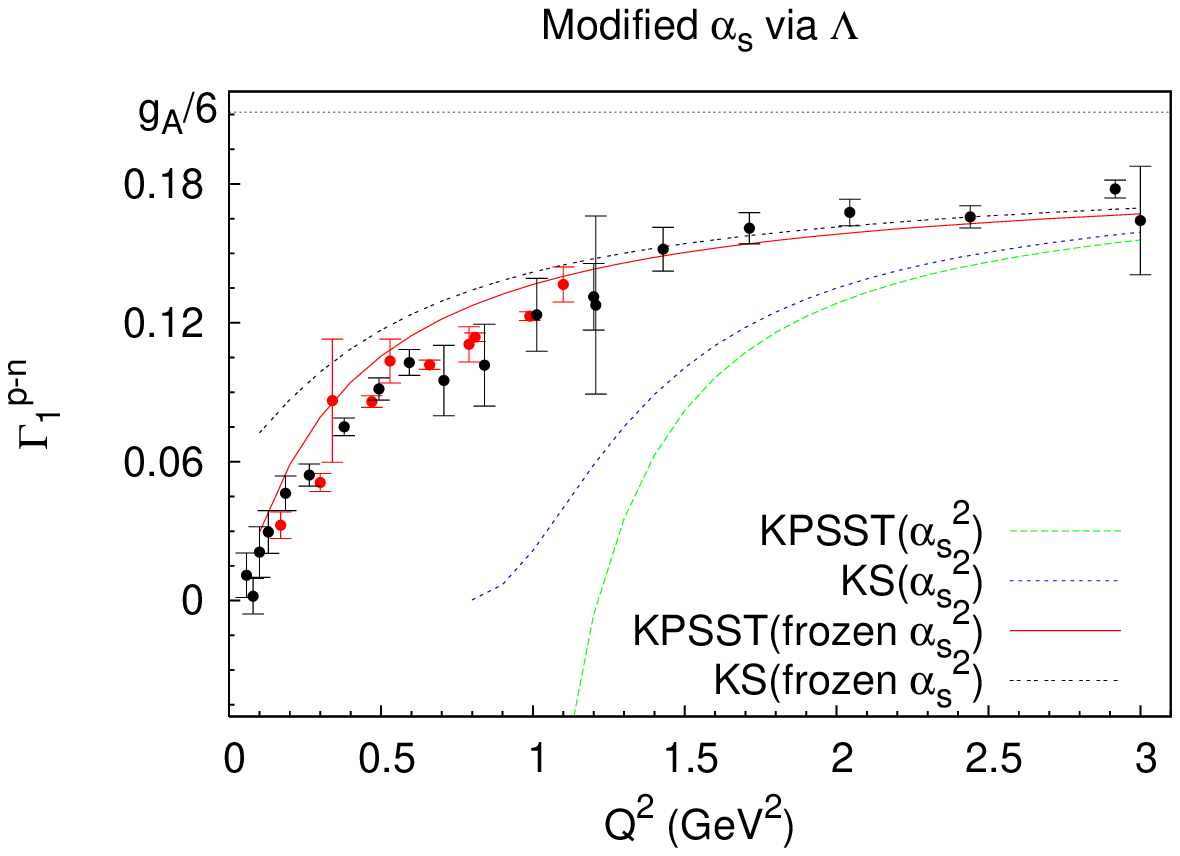}
\vskip -0.3cm
\caption{(color online). As in Fig 2 but $a=2$.}
\label{fig:F3}
\end{figure}

\section{Conclusion}
The analysis of the Bjorken sum rule performed within the framework of perturbative QCD
is presented at low $Q^2$.
It features
the following important steps:
\begin{itemize}
\item
The new form (\ref{2}) for the twist-two part was used. It is
compatible with
the observation
$E_{NS}(Q^2\to 0) \to 0$, coming from the experimental data
(if HTs are negligible).
\item
The application of the Grunberg method of effective charges
\cite{Grunberg:1980ja} in a combination
with a ``frozen'' coupling constant provides good agreement with
experimental data,
though with a slightly larger freezing parameter ($1.5 M^2_{\rho}$ instead of $M^2_{\rho}$).
\footnote{It seems that
the value of the parameter $a$
depends on the
order of perturbation theory. We plan to study this dependence in our forthcoming investigations.}
 \end{itemize}

Further elaborations to be undertaken include taking into account the $\alpha_s^2$ and $\alpha_s^3$
corrections to our analysis, as well as the study of HT corrections
and their correlations with a freezing parameter $a$ (in front
of $M^2_{\rho}$). We also plan to add to our analysis
an analytic coupling constant \cite{Shirkov:1997wi},
which has no the Landau pole and leads usually to the results, which are similar
to those obtained in the case of the ``frozen'' coupling constant~\cite{Kotikov:2004uf,Cvetic:2009kw}.\\

A.V.K. thanks the Organizing Committee of 20th International Symposium on
Spin Physics (SPIN2012) for invitation and support.


\end{document}